\documentclass[reprint,aps,prb,superscriptaddress,floatfix]{revtex4-2}
\usepackage{amsmath}
\usepackage{amssymb}
\usepackage{mdsymbol}
\usepackage{graphicx, subfig}
\usepackage{dsfont}
\usepackage{braket}
\usepackage{hyperref}
\usepackage{amsfonts}

\usepackage{textcomp}
\usepackage{times}
\usepackage{float}
\usepackage{latexsym,amsmath,amssymb,bm,euscript}
\usepackage{color}
\usepackage{epstopdf}
\usepackage{soul}
\usepackage[normalem]{ulem}
\usepackage{xspace}
\usepackage{natbib}
\usepackage{ulem}
\usepackage{etoolbox}
\usepackage{lipsum} 

\newcommand{\wlx}[1]{{\color[rgb]{.8,.8,.8}{\sout{#1}}}}

\begin{document}
\title{Phase diagram of a one-dimensional Ising model with an anomalous $\mathbb Z_2$ symmetry}
\author{Jin-Xiang Hao}
\affiliation{State Key Laboratory of Surface Physics and Department of Physics, Fudan University, Shanghai 200433, China}

\author{Wei Li}
\email{w.li@itp.ac.cn}
\affiliation{CAS Key Laboratory of Theoretical Physics, 
Institute of Theoretical Physics, Chinese Academy of 
Sciences, Beijing 100190, China}
\affiliation{CAS Center of Excellence in Topological Quantum Computation, 
University of Chinese Academy of Sciences, Beijing 100190, China}

\author{Yang Qi}
\email{qiyang@fudan.edu.cn}
\affiliation{State Key Laboratory of Surface Physics and Department of Physics, Fudan University, Shanghai 200433, China}
\affiliation{Collaborative Innovation Center of Advanced Microstructures, Nanjing 210093, China}
\begin{abstract}
Anomalous global symmetries, which can be realized on the boundary of symmetry-protected topological phases, bring\wlx{s} 
new phases and phase transitions to condensed matter physics. In this work, we study a one-dimensional model with an
anomalous $\mathbb Z_2$ symmetry, using the density-matrix renormalization group method. Besides a symmetry-breaking 
ferromagnetic phase, we find a gapless phase described by the Tomonaga-Luttinger Liquid theory with the central charge $c=1$.
Our numerical finding is compatible of theoretical constraints on possible phases resulting from the symmetry anomaly.
\end{abstract}
\maketitle

\section{Introduction}

Symmetry is an essential ingredient in the study of condensed matter physics.
In traditional Landau paradigm~\cite{landau1937theorie},
symmetries classify phases, and the spontaneous breaking of symmetries describes continuous phase transitions.
In the last few decades, the discovery of topological states of matter, such as the intrinsic
\cite{WenNiu1990TO,Wen1990TO} and symmetry-protected topological (SPT) orders~\cite{Gu2009,ChenCZX,ChenSPTScience,ChenSPTPRB} (including topological insulators and topological superconductors~\cite{Hasan2010,Qi2011}), greatly expands 
our understanding of phases and phase transitions. In particular, on the boundary of an SPT state,
the symmetries are anomalous~\cite{Wen2013Anomaly}, meaning that the microscopic realization of the symmetries must be nonlocal.
Such anomalous global symmetries can also be realized in a standalone system instead of on a boundary, if only the low-energy degrees of freedom (like half-integer spins) are considered~\cite{Meng2016LSM}.
Phases and phase transitions with anomalous global symmetries are different from their counterparts with anomaly-free symmetries.
Most importantly, anomalous symmetries do not allow phases that are gapped, symmetric and trivial (without any intrinsic topological order).
In general, anomalous symmetries lead to new phases, e.g., nontrivial gapless phases and phase transitions. 

Although anomalous symmetries are generally realized as the boundary of a topological phase in one-higher dimension,
numerical simulations using this realization is challenging because of the extra computational cost of simulating the 
bulk~\cite{YCWang2017QSL}. Therefore, it is desirable to study anomalous models without a bulk.
Because of the topological nature of the anomaly, such models must be nonlocal: they either have a nonlocal 
Hamiltonian, or a nonlocal symmetry action. For the former approach, anomalous fermionic states violating the 
fermion-doubling theorem can be realized by the so-called SLAC-fermion 
Hamiltonian~\cite{*[] [{. The model is named after the place it was invented: the Stanford Linear Accelerate Center (SLAC).}] SLACFermion}, which contains long-range hopping terms.
For the latter approach, \citet{chatterjee2022algebra} introduces systematic ways to construct such nonlocal symmetry actions realizing certain anomalies,
and such models with anomalous $\mathbb{Z}_{2}\times\mathbb{Z}_{2}$
symmetry and $S_3$ symmetry are used to simulate phases and phase transitions with anomalous symmetries without the bulk~\cite{chatterjee2022holographic}.
Comparing to the former approach, these models are easier to handle numerically because the Hamiltonian is still local.
Therefore, this is a promising way to study phase and phase transitions with anomalous symmetries.

In this work, we study a one-dimensional (1d) spin model introduced in Ref.~\cite{chatterjee2022algebra} with an anomalous $\mathbb Z_2$ symmetry numerically, using the density-matrix renormalization group (DMRG) method.
The anomaly indicates it is the boundary of a two-dimensional (2d) $\mathbb Z_2$ SPT state.
Therefore, it does not have a trivial paramagnetic phase, which is gapped and symmetric.
Through tuning the coupling constants, we indeed find that the model realizes two phases: a symmetry-breaking ferromagnetic (FM) phase, which also appears in models with anomaly-free $\mathbb Z_2$ symmetry, and a gapless phase with a central charge $c=1$, which is described by the Tomonaga-Luttinger Liquid (TLL) theory.
In fact, at an exactly-solvable line in the gapless phase, the model can be mapped to a 1d free-fermion system, which corresponds to a TLL with the Luttinger parameter $K=1$.
The rest of the gapless phase is also a TLL with $K>1$.


The rest of the paper is organized as follows. In Sec.~\ref{sec:model}, we introduce the 1d model with an anomalous $\mathbb Z_2$ symmetry proposed in Refs.~\cite{ji2020categorical,chatterjee2022algebra}, and discuss possible phases compatible with the anomaly, using the boundary-bulk correspondence.
In Sec.~\ref{sec:numerical}, we give details of the DMRG method
we use, and the physical quantities we compute. Section~\ref{sec:results} describes the numerical results we obtained, and   Sec.~\ref{sec:discuss} is devoted to the discussion.

\section{Model}
\label{sec:model}

In this work, we study the 1d lattice model introduced in Ref.~\cite{chatterjee2022algebra}, which has an anomalous $\mathbb Z_2$ symmetry.
The model Hamiltonian is the following,
\begin{equation}
  \label{201646}
  \begin{aligned}
  H=&-J_z\sum_{i=1}^{L}Z_iZ_{i+1}-h_1\sum_{i=1}^{L}\left(X_i-Z_{i-1}X_iZ_{i+1}\right)\\
  &-h_2\sum_{i=1}^{L}Z_{i-1}\left(X_i+Z_{i-1}X_iZ_{i+1}\right),
  \end{aligned}
\end{equation}
where $L$ is the number of sites, and $X_i$ ($Z_i$) is Pauli  X(Z) operator on the site $i$.

The Hamiltonian \eqref{201646} has an anomalous $\mathbb{Z}_2$ symmetry~\cite{chatterjee2022algebra},
described by the symmetry operator
\begin{equation}\label{202226}
  W=\prod_{i}X_i\prod_{i}{\rm i}^{\frac{-Z_i+Z_{i+1}+Z_iZ_{i+1}-1}{2}}.
\end{equation}
This is an anomalous non-on-site symmetry because it cannot be expressed as a product of local operators.
In addition to flipping spins, the $W$ operator adds a phase factor for each nearest-neighbor pair of spins:
the phase factor is $+1$ if the spins are $\ket{\uparrow\uparrow}$, $\ket{\downarrow\downarrow}$ or
$\ket{\downarrow\uparrow}$,
and $-1$ if the spins are $\ket{\uparrow\downarrow}$.
Therefore, the overall action of $W$ is flipping all spins and adding a $\pm1$ phase if the total number of $\ket{\uparrow\downarrow}$ domain walls is even (odd), respectively.
Since only the $\ket{\uparrow\downarrow}$ domain walls are counted, $W$ only strictly satisfies the $\mathbb Z_2$ algebra $W^2=\mathds1$ if the boundary condition is periodic.
However, the slight violation of $W^2=\mathds1$ by other boundary conditions should have negligible effect in the thermodynamic limit, which is also confirmed by our DMRG simulation.
Therefore, in the DMRG simulation, we still mostly use open boundary conditions because it reduces the bond dimensions needed in the simulation.

Two possible phases in this 1d model can be derived using the anomalous $\mathbb Z_2$ symmetry and the boundary-bulk correspondence of SPT states.
Because of the anomaly, this 1d model can be realized on the edge of a 2d nontrivial SPT state protected by the $\mathbb Z_2$ symmetry, which is known as the Levin-Gu~\cite{LevinGu2012} or CZX state~\cite{*[] [{. The SPT state described in this work is called the CZX state, because it is constructed using the control-X (CZX) quantum gate.}] ChenCZX}.
An essential feature of the 2d SPT state is that its boundary is either gapless or spontaneously breaks the $\mathbb Z_2$ symmetry.
Therefore, the 1d model also has two possible phases: a symmetry-breaking FM phase, and a symmetric gapless phase.

In deed, these two possible phases are observed in model \eqref{201646} along the line of $h_2=0$, where the model can be solved exactly using the Jordan-Wigner transformation.
It is well-known that the 1d transverse-field Ising model can be solved exactly with this method, and the key to the transfoermation is the ordinary onsite $\mathbb Z_2$ symmetry generated by $W_0=\prod_iX_i$, which is mapped to the fermion-parity symmetry by the Jordan-Wigner transformation.
It is straightforward to check that the model \eqref{201646} also has this onsite $\mathbb Z_2$ symmetry $W_0$ (in addition to the anomalous $\mathbb Z_2$ symmetry $W$) when $h_2=0$, and hence can be studied by the same transformation, which takes the following form:
\begin{equation}
  \label{eq:JW}
  \begin{split}
    Z_i &= 2c_i^\dagger c_i-1;\\
    X_i &= \prod_{j<i}\left(1-2c_j^\dagger c_j\right)\left(c_i^\dagger+c_i\right).
  \end{split}
\end{equation}
As discussed in details in Appendix~\ref{app:exactly solvable model},
this Jordan-Wigner transformation maps the original spin model to the following free-fermion model,
\begin{equation}
  \label{eq:ffmodel}
  \begin{split}
  H=&-J_z\sum_i\left(c_i^\dagger-c_i\right)\left(c_{i+1}^\dagger +c_{i+1}\right)\\
    &-h_1\sum_i\left[2c_i^\dagger c_i-1+\left(c_{i-1}^\dagger-c_i\right)\left(c_{i+1}^\dagger+c_{i+1}\right)\right].
  \end{split}
\end{equation}
The quasiparticle spectrum of this model is $E_k=\pm(J_z+2h_1\cos k)$.
When $h_1/J_z>\frac12$, the model is gapless because there are two Dirac cones in the quasiparticle dispersion at
\begin{equation}
  \label{eq:Dirac}
  k=\pm\cos^{-1}\left(-\frac{J_z}{2h_1}\right).
\end{equation}
Therefore, the model is discribed by gapless fermions, which is in turn mapped to a TLL with $c=1$ and $K=1$ using the bosonization technique.
(Here, the two Dirac cones appear in the quasiparticle dispersion of a Bogoliubov-de Geens Hamiltonian, and therefore each contributions $c=\frac12$.)
On the other hand, when $h_1/J_z<\frac12$, the free-fermion model is gapped and belongs to a nontrivial 1d topological superconductor.
This corresponds to an FM state in the original spin model.

In Sec.~\ref{sec:results}, we study the model at parameters $h_2>0$, where the model can no longer be mapped to free-fremion models because of the absence of the onsite $\mathbb Z_2$ symmetry $W_0$.
If the gapless phase is smoothly extended to regions of the phase diagram with $h_2>0$, we expect it to remain in the TLL phase with $c=1$.
However, as the gapless phase may deviate from free-fermion, the Luttinger parameter $K$ may differ from 1.
In fact, it is found to be the case in our numerical simulation.  

We notice that a TLL with $c=1$ is only one of possible gapless edge states.
In fact, using recent progresses in the correspondence between gapless edges and bulk topological orders~\cite{Kong2020,*Kong2021} and between anomalous edge CFT and bulk SPTs~\cite{Scaffidi2016,Meng2020RA},
one can see that not only the TLL, but infinitely many other gapless states can be realized on the boundary of double-semion topological order or $\mathbb Z_2$-SPT state.
Among them, the TLL is the simplest and most likely to be realized in the lattice model \eqref{201646}.

\section{Numerical methods}
\label{sec:numerical}

We employ the DMRG method to simulate the model Eq. (\ref{201646}), with the ITensor library~\cite{itensor}.
In the calculations, we set a maximum bond-dimension of 2000 that guarantees very well converged results,
with truncation errors below $10^{-10}$.

        The order parameter of the $\mathbb Z_2$ symmetry in this
        model is the average magnetization per site $m=\frac1L\sum_iZ_i$, where $Z_i$ is Pauli  Z operator on the site $i$ (this can still be used as the order parameter despite of the anomaly, because of the form of the ``patch charge operator'' in this model~\cite{chatterjee2022algebra}).
        We use the binder ratio $U$,
        \begin{equation}\label{161726}
            U=1-\frac{\langle m^4\rangle}{3\langle m^2\rangle^2}
        \end{equation}
        to locate the phase transition point.

        During our calculation, we find that if we set the parameter $h_2$ to a relatively large value,
        the model will always be in the FM phase.
        On the other hand, when the parameter $h_2=0$, the model becomes exactly solvable as discussed in Sec.~\ref{sec:model} and in Appendix~\ref{app:exactly solvable model}.
        In addition, Hamiltonian \eqref{202226} with $h_2=0$ has the special property that the classical FM state is always an exact eigenstate.
        Consequently, this state will remain to be the exact ground state for $h_1$ smaller than a critical value, and this special property may affect the phase transition.
        Thus, we set the parameter to a small but nonvanishing value $h_2=0.3$ (while fixing $J_z=1$) in the most parts of Sec.~\ref{sec:results}.

            To determine what type of the CFT this model belongs to, we need to compute the central
        charge, which is the main characteristic of CFT. We can obtain it from the entanglement entropy of the model.
        For a one dimensional quantum system with a length $L$ and a peroidic boundary condition, the entanglement entropy $S_E$ should
        follow a linear scaling with the conformal distance $\tilde{l}$~\cite{calabrese2004entanglement},
        \begin{equation}\label{171626}
        \begin{aligned}
            S_{E}&=\frac{c}{3}\cdot{\rm ln}\left[\frac{4(L+1)}{{\rm\pi}}{\rm sin}\frac{{\rm \pi\left(2r+1\right)}}{2\left(L+1\right)}\right] + const.\\
                &=\frac{c}{3}\cdot\tilde{l}+const.,
        \end{aligned}
        \end{equation}
        where $r$ is the subsystem size and $c$ is the central charge. 
        We notice that the periodic boundary condition is often used for computing the central charge because it gives much better finite-size scaling behaviors than the open boundary condition, despite of the additional computational cost.

    \section{Results}
    \label{sec:results}
\begin{figure}[!tbp]
            \includegraphics[width=.4\textwidth]{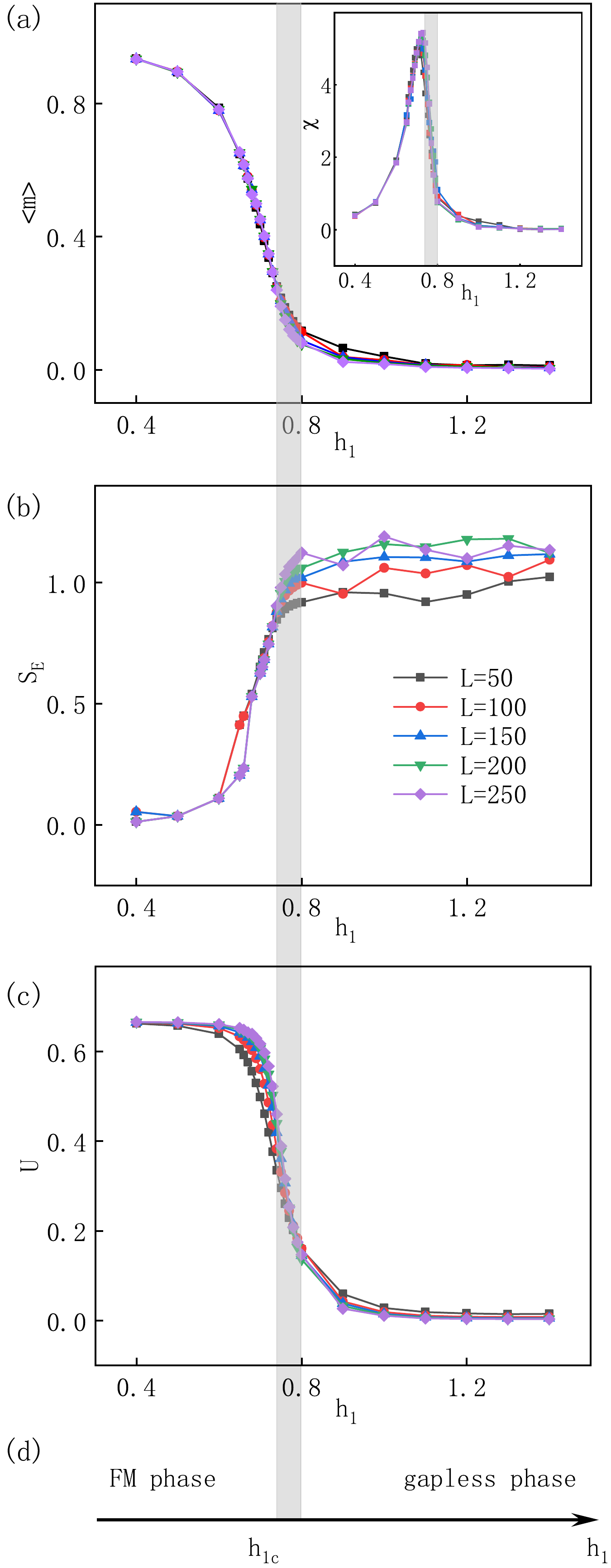}
            \caption{(a)~DMRG data for the order parameter (magnetization $\langle m\rangle$) versus $h_1$ for $J_z=1.0$ and
                    $h_2=0.3$, with system sizes $L=50, L=100, L=150, L=200$ and $L=250$ from 
                    top to bottom. The inset shows the corresponding $\chi$ as a function of $h_1$. 
                    (b)~The entanglement entropy $S_{E}$ for the half subsystem size versus $h_1$
                    in the same condition. (c)~The binder ratio $U$ of the system
                    as a function of $h_1$ in the same condition. The curves at different system sizes cross approximately at the critical $h_c$.
                    We calculate (a), (b), and (c) at open
                    boundary condition for efficiency, which is no effect on the results. (d)~The phase diagram of the system for $h_1$. The grey line across all panels indicates the approximate localtion of the phase transition.}
            \label{162246}
        \end{figure}

            As a first step, we draw the phase diagram of the model Eq.~(\ref{201646}) from DMRG data. To this end,
        we fix $J_z=1$ and $h_2=0.3$, and calculate the order parameter $\langle m\rangle$, the susceptibility $\chi=\frac{\partial\langle m\rangle}{\partial h_1}$, and the binder ratio [see Eq. (\ref{161726})] of the
        1d quantum system as a function of the parameter $h_1$ at different system sizes, as shown in Fig.~\ref{162246}.
        Here, we get the results in open boundary condition for speeding up the calculation, and it has no effect on the results. Fig.~\ref{162246}d shows the phase diagram of the system.
        
        From Fig.~\ref{162246}a, we can see that the order parameter $\langle m\rangle$
        goes from finite values to zero as increasing the value of $h_1$. This indicates that the system undergoes
        a phase transition from an FM phase to a $\mathbb Z_2$-symmetric phase as increasing $h_1$.
        In fact, the FM phase with small $h_1$ is smoothly connected to the point of $h_1=h_2=0$, where the model reduces to the classical ferromagnetic Ising model.
        Next, we locate the phase transition point at $h_{1c}\approx 0.76(4)$ from these measurements.
        First, the susceptibility $\chi$ shows a sharp peak at $h_1=0.72$~\footnote{The peak is not divergent as $L$ increases, and this is consistent with the phase transition being of Berezinskii-Kosterlitz-Thouless type, which will be discussed later.}.
        Additionally, Fig.~\ref{162246}c shows that the binder ratios of different system sizes also cross at $h_1=0.80$, and the smooth crossing indicates that the phase transition is continuous.
        Determining the precise location of binder-ratio crossing turns out to be difficult, due to the oscillating finite-size dependence explained in Appendix~\ref{app:finite-size}.
        We suspect that this oscillation is related to the fact that the gapless fermionic mode locates at nonzero momenta in Eq.~\eqref{eq:Dirac}.
        Comparing these two measurements, we put the transition point between $0.72$ and $0.80$, and leave the task of finding the  precise location of the quantum critical point to future works.
        
\begin{figure*}[htbp]
            \centering
            \subfloat[]{
                \label{06161603}
                \includegraphics[width=.4\textwidth]{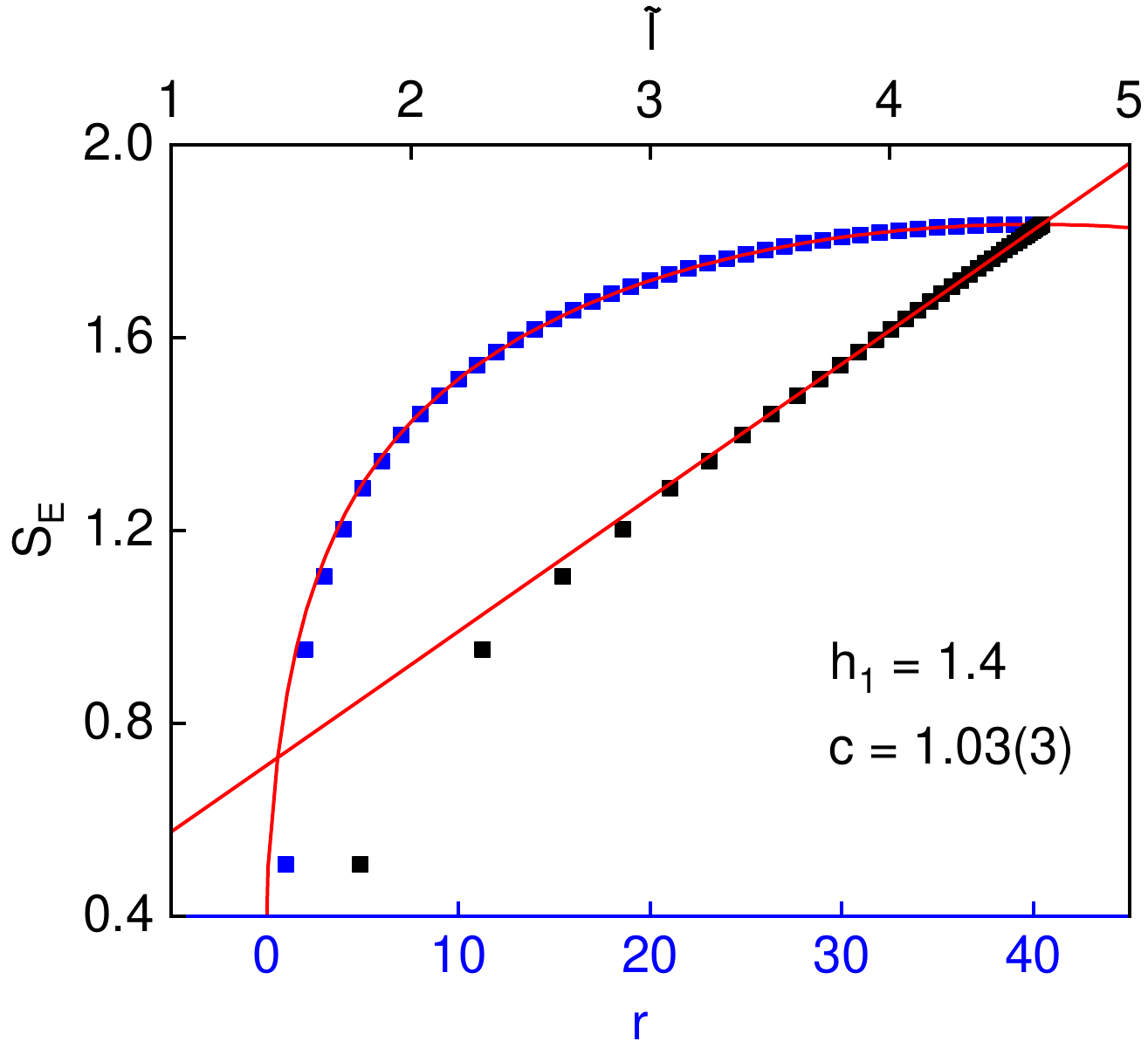}
                }
            \subfloat[]{
                \label{06161607}
                \includegraphics[width=.4\textwidth]{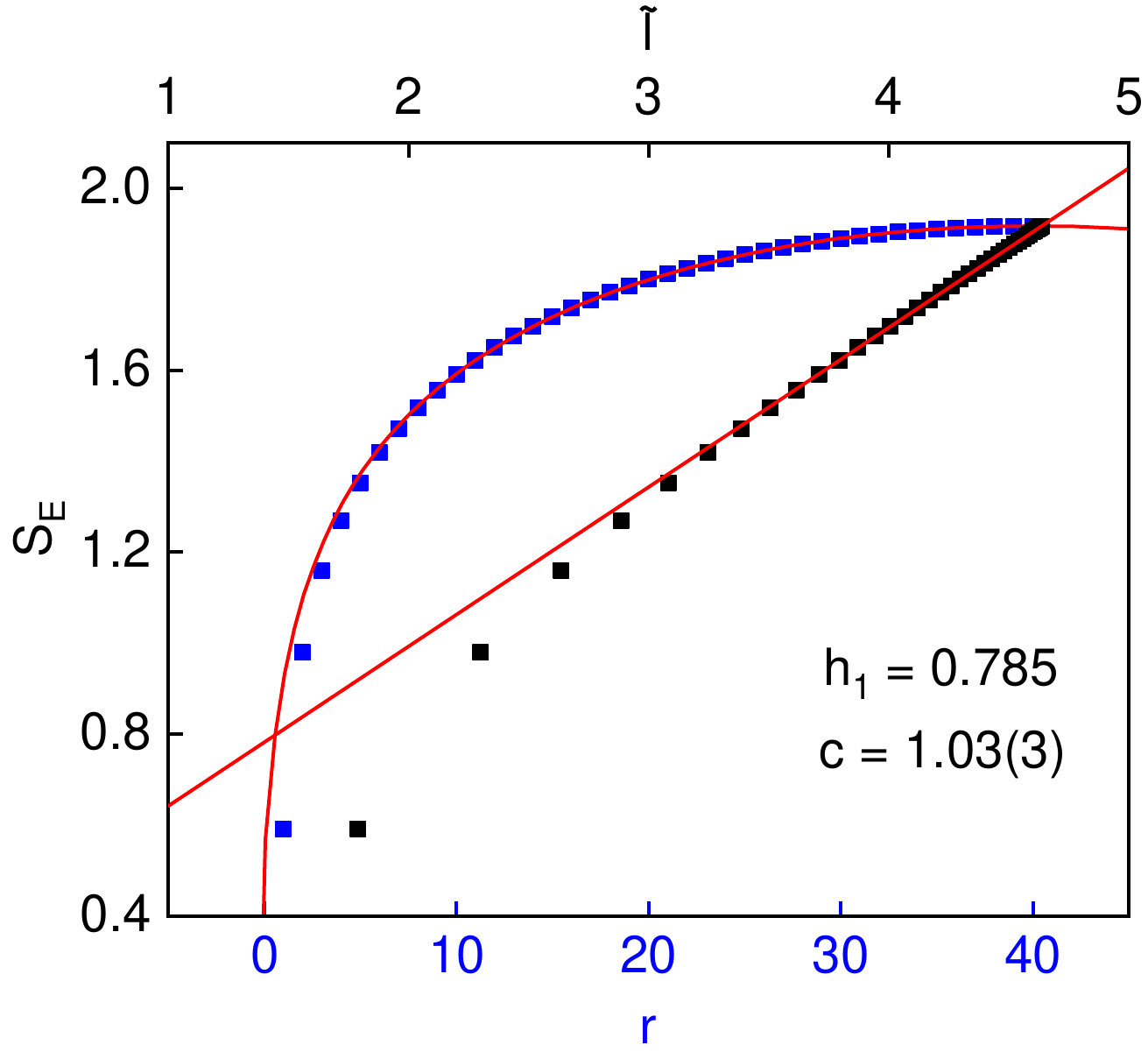}
                }\\
            \subfloat[]{
                \label{06262121}
                \includegraphics[width=.42\textwidth]{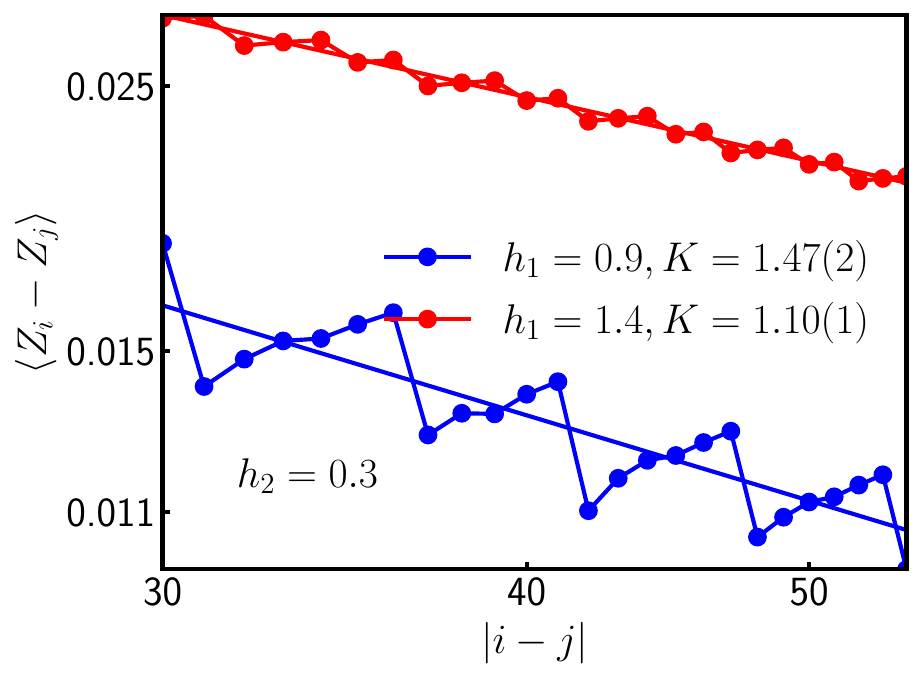}
                }
            \subfloat[]{
                \label{06262122}
                \includegraphics[width=.4\textwidth]{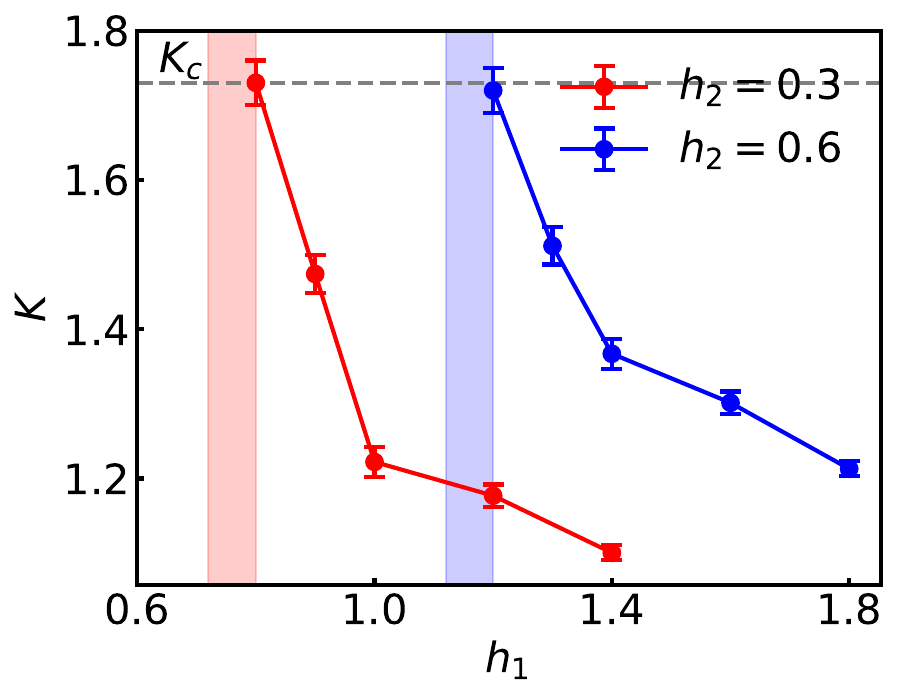}
                }
            \caption{\protect\subref{06161603}~The blue data show that the entanglement entropy $S_E$ of the system versus the subsystem size r for $J_z=1.0$,
            $h_1=1.4$, $h_2=0.3$, and the system size $L=80$.
            The black data show that the variation of entanglement entropy $S_E$ of the
            system versus the conformal distance $\tilde{l}$ for $J_z=1.0$, $h_1=1.4$, $h_2=0.3$, and the system size $L=80$. After discarding
            the first eight data due to the boundary effect, we can get the central charge $c=1$ by fitting the slope of the curve.
            We calculated the results in the periodic boundary condition. \protect\subref{06161607}~The entanglement entropy of the system for
            $h_1=0.785$, near the critical point, in the same condition. 
            \protect\subref{06262121}~The Luttinger parameter $K$ with two different $h_1$ values. In the condition $h_2=0.3$, with $h_1=0.9$ and $h_1=1.4$, we get the Luttinger parameter $K=1.47(2)$ and $K=1.10(1)$, respectively. 
            We obtained $K$ from the slope of the data for variation of the correlation function $\langle Z_iZ_j\rangle$ at site distance $|i-j|=L/2$ versus the system size $L$ on a log-log scale. We calculated it at periodic boundary condition to diminish the oscillations. 
            \protect\subref{06262122}~Detailed study of $h_1$ dependence of the Luttinger parameter $K$ with two different $h_2$ values: $h_2=0.3$ and $h_2=0.6$. The shaded regions in the figure indicate the range of the critical points for the corresponding $h_2$ values. Note that the finite-size oscillations of the correlation function become increasingly pronounced as the critical point is approached, causing the error bars of the extracted $K$ to gradually increase near criticality.
            Moreover, for $h_2 = 0.3$ and $h_2 = 0.6$, the critical points are located near $h_1 = 0.80$ and $h_1 = 1.2$, respectively. At these $h_1$ values, the Luttinger parameter takes $K = 1.73(3)$ and $K = 1.72(3)$. 
            The closeness of these two values indicates that the CFT parameter $K$ at the critical points is the same for different $h_2$, consistent with the nature of the BKT phase transition.}
            \label{172314}
        \end{figure*}

            In order to determine the nature of the phase with zero magnetization, we calculate the entanglement entropy $S_{E}$ for the half subsystem size as a function 
        of $h_1$ at different system sizes, as shown in Fig.~\ref{162246}b. We can see that below the critical point $h_{1c}$, the value of $S_E$ is zero, and 
        beyond $h_{1c}$, the value of $S_E$ varies with the system size $L$. Moreover, we compute the entanglement entropy $S_{E}$ as a function of the
        subsystem size r for $J_z=1.0$, $h_1=1.4$, $h_2=0.3$, and a fixed system size $L=80$ with the periodic boundary condition, and the results are shown in
        Fig. \ref{06161603} (blue data). We find that the value of the entanglement entropy increases with the increase of $r$. These results indicate that the
        phase, with a vanishing order parameter for $h_1>h_{1c}$, is a nontrivial gapless phase. This gapless phase can be
        described by a 1d CFT.

            To determine the type of the CFT for the gapless phase, we compute the central charge of the CFT.
        According to Eq. (\ref{171626}), we plot the entanglement entropy $S_{E}$ as a function of the
        conformal distance $\tilde{l}$ (shown in Fig.~\ref{06161603}) to obtain the central charge of the system from the slope of the function.
        We can see that except for the first few data points due to the boundary
        effect, the expected linear dependence for the entanglement entropy is obtained as the conformal distance
        increases, and we extract the central charge $c=1$ from the slope of the linear dependence. Moreover, we calculate the entanglement
        entropy and the central charge of the system near the critical point (see Fig.~\ref{06161607}). The results indicate that the system
        at the critical point $h_{1c}$ is also a $c=1$ CFT.

            It is well known that a large class of CFTs with central charge $c=1$ is described by Luttinger liquids \cite{luttinger1960fermi}. The Luttinger parameter
        is the main characteristic of Luttinger liquids, which we can extract from the power-law scaling of the correlation function
        $\langle Z_iZ_j\rangle$,
        \begin{equation}\label{182227}
            \langle Z_iZ_j\rangle=\left|i-j\right|^{-K/2},
        \end{equation}
        where $K$ is the Luttinger parameter.

        We now turn to the measurement of the Luttinger parameter of the system.
        As shown in Fig.~\ref{06262121}, we plot, on a log-log scale,
        the correlation function $\langle Z_iZ_j\rangle$ as a function of the
        site distance $\left|i-j\right|$ for $h_2=0.3$ in a periodic boundary condition, with two values of $h_1$: $h_1=0.9$ near the critical point and $h_1=1.4$ in the gapless phase.
        In the plot, the correlation functions exhibit an overall power-law-decay behavior.
        In addition, they also oscillate with distance, indicating that the critical $\langle Z_iZ_j\rangle$ correlation occurs at a finite momentum, which is also consistent with a similar behavior at the exactly solvable point.
        From the correlation functions, we can extract the values of Luttinger parameter $K$ from the slope of the log-log plot according to Eq.~\ref{182227}.
        
        In the gapless phase, we observe that when $h_2>0$, the system deviates from the exact solution, and $K$ grows larger than 1.
        As shown in Fig.~\ref{06262121}, at $h_2=0.3$, $h_1=0.9$ and 1.4, we get $K=1.10$ and 1.47, respectively.
        Moreover, Fig.~\ref{06262122} shows several measurements of $K>1$ at different parameters in the gapless phase.
        These show that the gapless phase is a Luttinger liquid, with variable Luttinger parameter $K>1$.

        We also carry out a prelimitary study on $K$ at the phase transition between the gapless phase and the FM phase.
        From measurements of $K$ as a function of $h_1$ for two different values of $h_2$: $h_2=0.3$ and $h_2=0.6$, as shown in Fig.~\ref{06262122},
        we can see that $K$ approaches the same value near the respective critical points~\footnote{For $h_2=0.6$, the critical point $h_{1c}\approx1.16(4)$ is determined from the Binder ratio $U$ as a function of $h_1$ with different system sizes.}. The agreement of these two values within error suggests that the CFTs at the critical points for different $h_2$ share the same value of $K$, which is consistent with the nature of the BKT phase transition.
        We leave a detailed study of this phase transition to future works.


\section{Discussion}
\label{sec:discuss}
In this work, we study the 1d model with an anomalous $\mathbb Z_2$ symmetry introduced in Refs.~\cite{ji2020categorical,chatterjee2022algebra}, using the DMRG method.
We find that the model undergoes a phase transition from a FM order to a gapless phase when varying the parameter $h_1$, and we locate the critical point from the binder ratio of the $\mathbb Z_2$ order parameter.
In addition, we extract the central charge $c=1$ from the entanglement entropy scaling, and obtain Luttinger parameter $K$ by fitting the exponent of the correlation function of the order parameter.
Both the absence of a trivial PM phase and the CFT of the gapless phase agrees with theoretical expectations due to the anomaly.


Our results not only reveals an interesting phase diagram of the anomalous 1d model and its intriguing resemblance of 1d spin-$\frac12$ chains,
but also demonstrates the power of studying models with anomalous global symmetries with a nonlocal symmetry action and without an SPT bulk.
We expect that systems with anomalous global symmetries, in both 1d and in higher dimensions, contain even richer phenomena than the traditional Landau-type symmetry-breaking phases and phase transitions, and models with nonlocal symmetry actions provide a viable way 
to study them.

\begin{acknowledgments}
This work was supported by National Natural Science Foundation of China (NSFC) through Grant Nos. 11874115, 12174068, 12222412, 11974036, 11834014, and 12047503.  
\end{acknowledgments}

\appendix

\section{Oscillating size dependence of binder ratio}
\label{app:finite-size}

\begin{figure*}[hbp]
    \subfloat[\label{fig:binder-zoomin}]{\includegraphics[width=.4\textwidth]{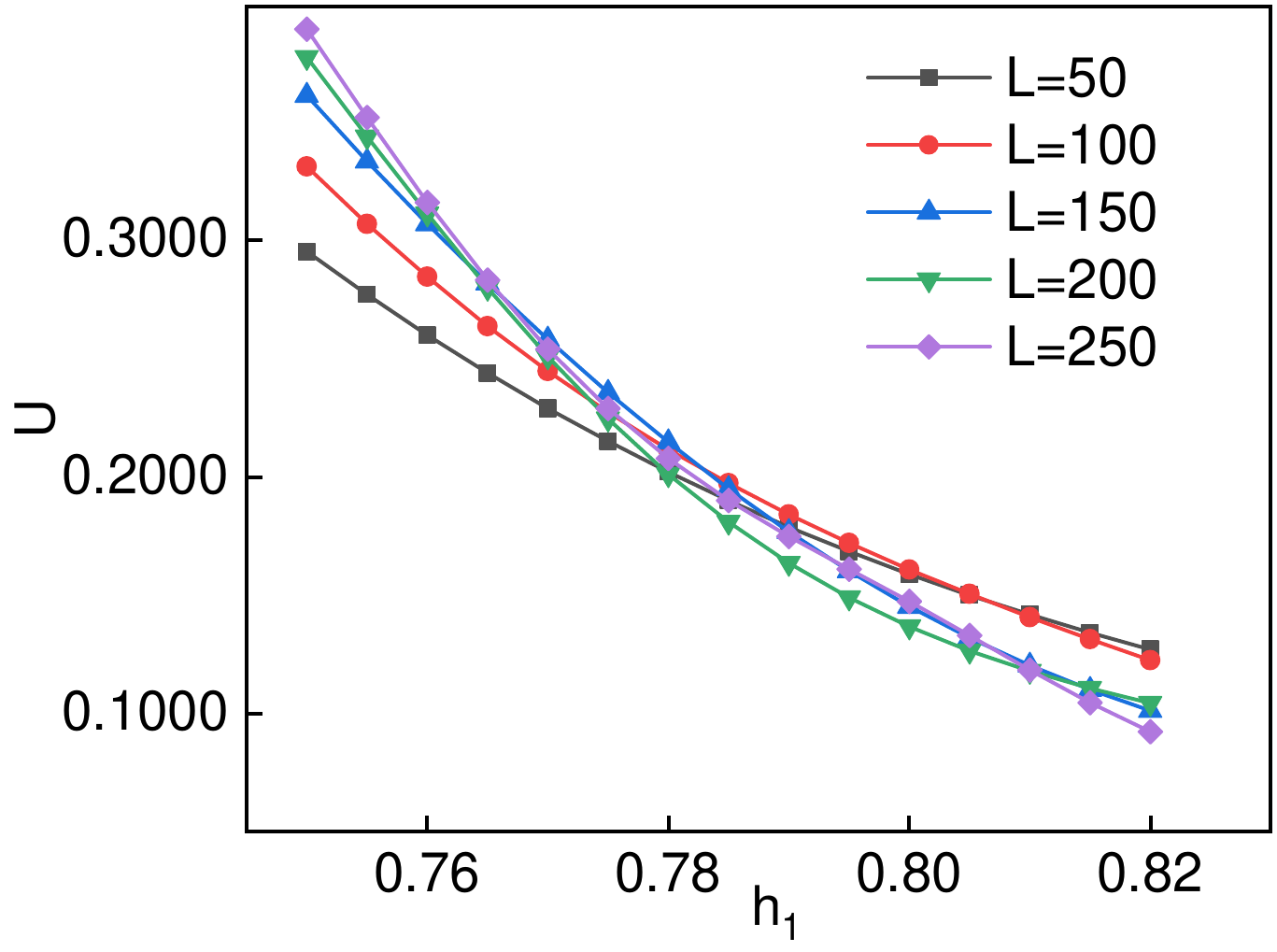}}
    \subfloat[\label{fig:binger-osc:critical}]{\includegraphics[width=.4\textwidth]{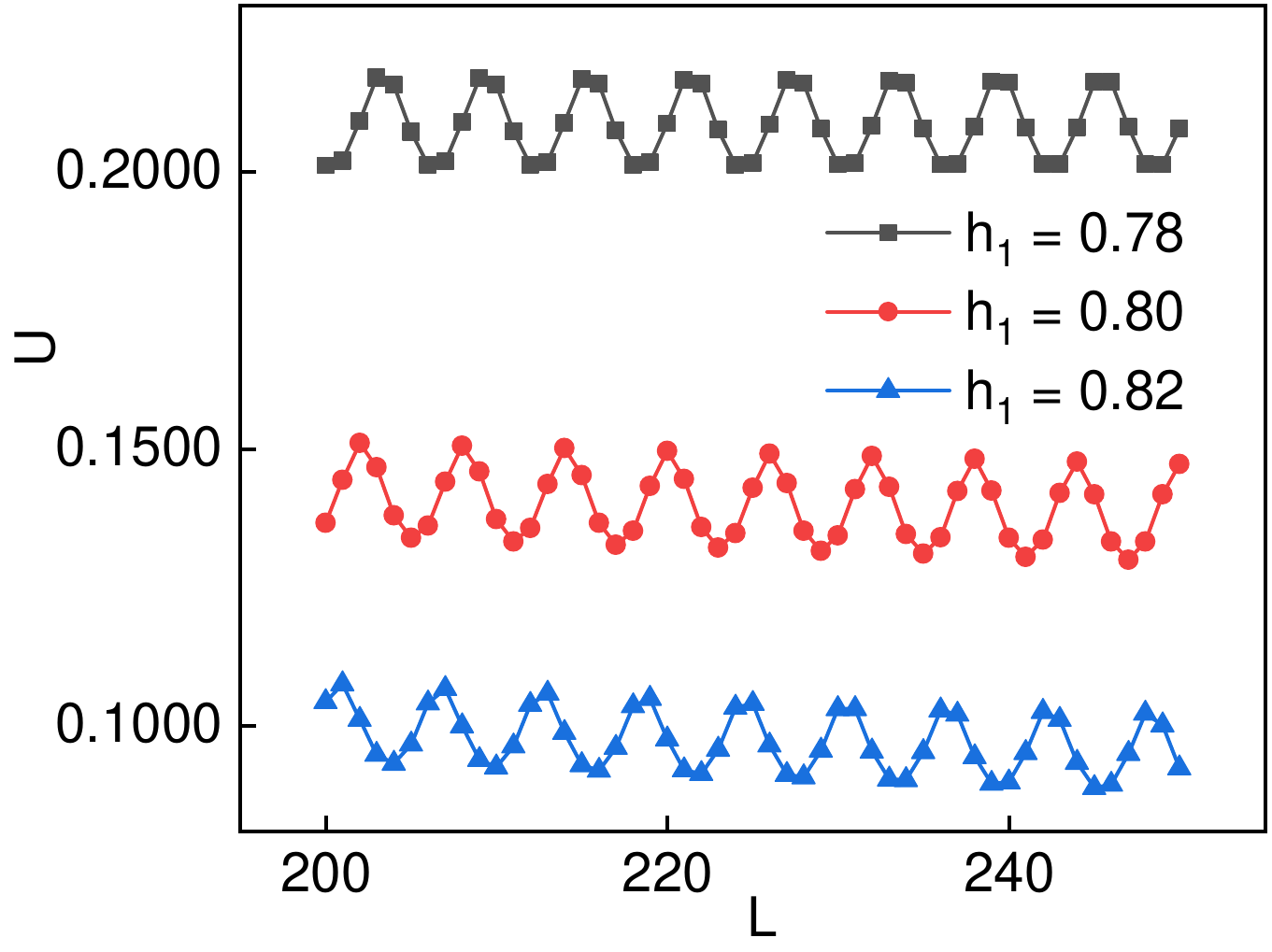}}\\
    \subfloat[\label{fig:binger-osc:gapless}]{\includegraphics[width=.4\textwidth]{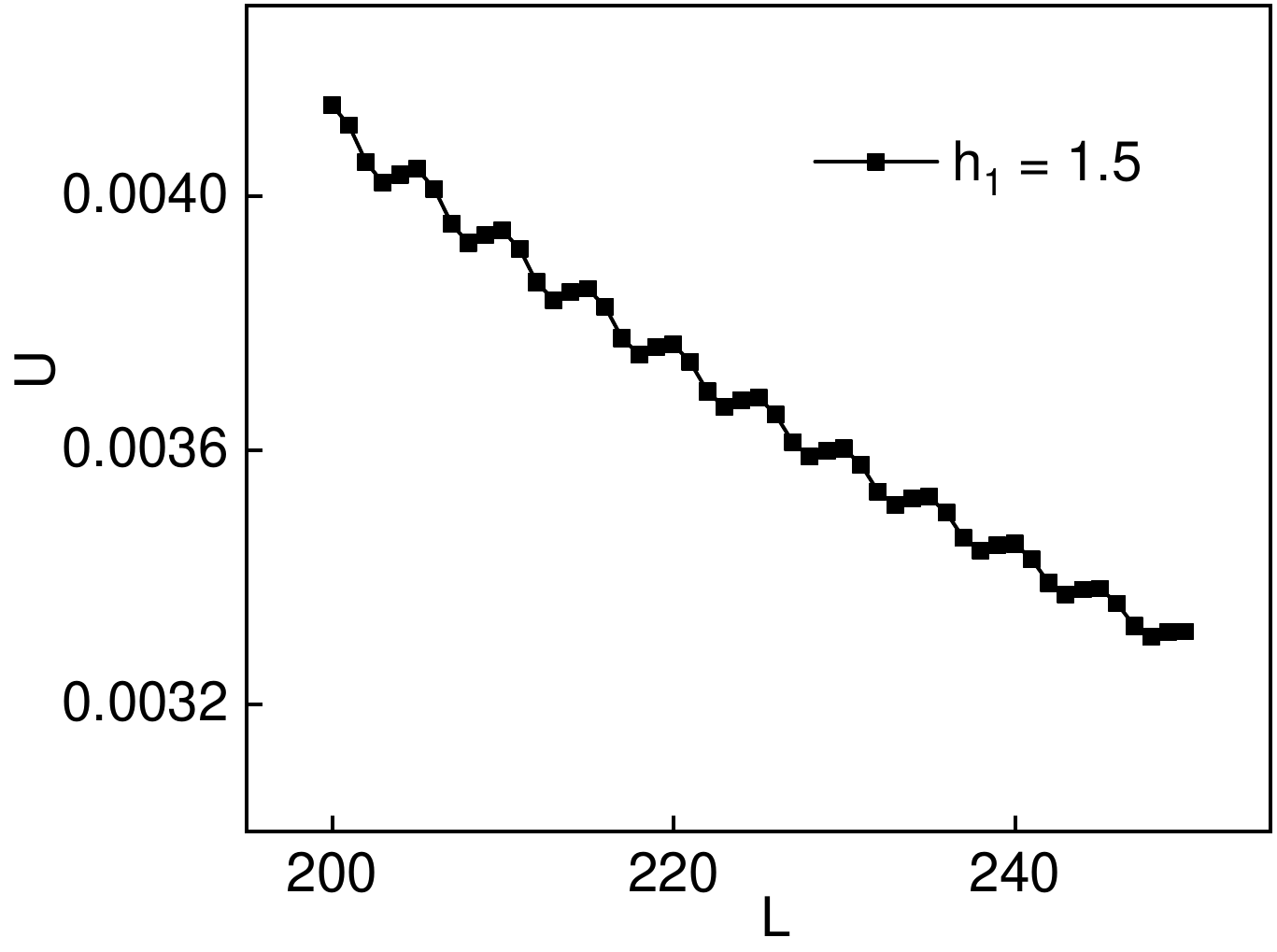}}
    \subfloat[\label{fig:binger-osc:fm}]{\includegraphics[width=.4\textwidth]{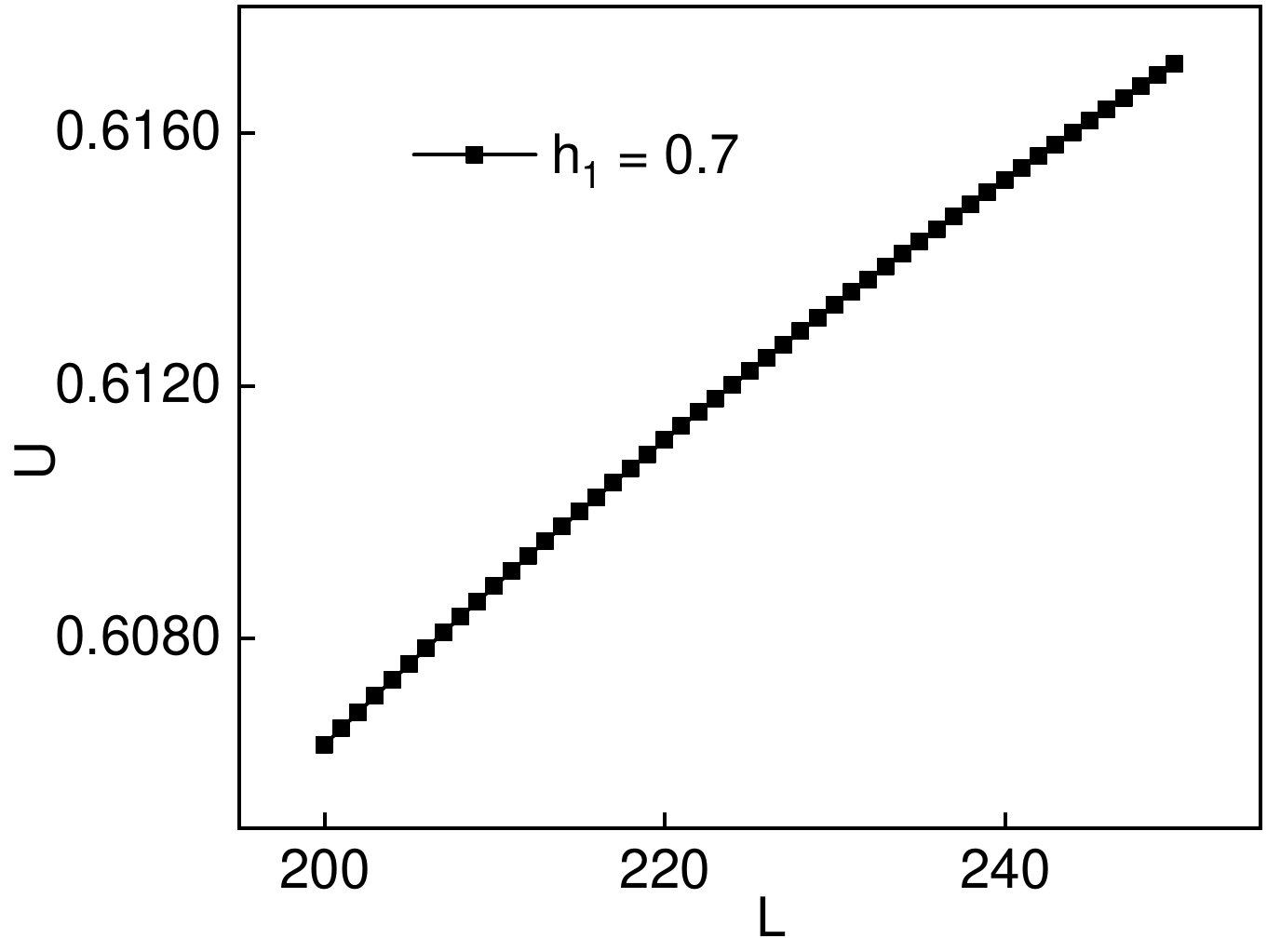}}
    \caption{\label{fig:binder-osc} Detailed study of $L$ and $h_1$ dependence of the binder ratio. (a) A zoom-in version of binder-ratio crossing shown in the main text. (b) Binder ratio as a function of system size, computed at values of $h_1$ near the critical point. (c) Binder ratio as a function of system size, computed at $h_1=1.5$, which is in the gapless phase. (d) Binder ratio as a function of system size, computed at $h_1=0.7$, which is in the FM phase.
    }   
\end{figure*}

In this section, we discuss an oscillating finite-size dependence of binder ratio. This behavior affects the determination of binder-ratio crossing, which is used in the main text to locate the phase-transition point $h_{1c}$.
In fact, although the binder ratio shows a crossing behavior in Fig.~\ref{162246}c in the main text, a zoom-in study of the plot in Fig.~\ref{fig:binder-zoomin} exhibits irregular $h_1$ dependences at different system sizes, and the curves do not cross at a single point.

To further investigate the size dependence of the binder ratio, in Fig.~\ref{fig:binder-osc}, we plot the system-size dependence of the binder ratio, at different values of $h_1$.
The results show that when $h_1\geq h_{1c}$, i.e. at the critical point or in the gapless phase, the binder ratio oscillates with $L$, with a period about 5-6.
In contrast, the binder ratio in the FM phase varies smoothly with $L$ without any oscillation.
This is consistent with the oscillating $\langle Z_iZ_j\rangle$ correlation function shown in Fig.~\ref{172314}\subref{06262121}-\subref{06262122}.
Therefore, we infer that the oscillating behavior in the binder ratio are caused by the finite-momentum spin correlation in the gapless phase:
Because of the dominant spin correlation in the gapless phase occurs at a finite momentum, all thermodynamic quantities, including the binder ratio, exhibits an oscillating behavior depending whether the system size is commensurate with the period of the spin correlation or not.
Furthermore, the oscillating behavior in Fig.~\ref{fig:binder-osc}b-c is responsible for the irregular $h_1$ dependence in Fig.~\ref{fig:binder-osc}a, because the binder ratio also fluctuates with $h_1$ at a fixed $L$ as the location of peak and valley in Fig.~\ref{fig:binger-osc:critical} also shifts with $h_c$.
Therefore, we will leave precise determination of phase transition point to future works.

\section{The model in exactly solvable case}
\label{app:exactly solvable model}

The Hamiltonian Eq.~(\ref{201646}) with $h_2=0$,
\begin{equation}
    \label{2301142040}
    \begin{aligned}
        H=-J_z\sum_{i=1}^{L}Z_iZ_{i+1}-h_1\sum_{i=1}^{L}\left(X_i-Z_{i-1}X_iZ_{i+1}\right),
    \end{aligned}
\end{equation}
is exactly solvable. In this appendix, we exactly obtain the eigenvalues of the Hamiltonian Eq.~(\ref{2301142040}) using the Jordan-Wigner transformation. 
From the dispersion relation of the model, we find that it has two Dirac cones with central charge $c=1$ and Luttinger parameter $K=1$.\par

Firstly, we briefly introduce the Jordan-Wigner transformation. It is well known that for a lattice model, $\frac{1}{2}$-spin basis on the site $i$ are $\ket{\uparrow}_i$ and $\ket{\downarrow}_i$, which can be converted each other by ladder operators $\sigma^{\pm}_i$,
\begin{equation}
    \label{2301122001}
    \begin{aligned}
        \sigma^{\pm}_i=\frac{1}{2}\left(X_i\pm{\rm i}Y_i\right).
    \end{aligned}
\end{equation}
Similarly, free fermion basis are also two level states $\ket{0}_i$ and $\ket{1}_i$, connected with each other by fermionic creation operator $c_i^{\dagger}$ and annihilation operator $c_i$. Therefore, one can map an interacting spin system 
to a free fermion system by the Jordan-Wigner transformation defined as,
\begin{equation}
    \label{2301121000}
    \begin{aligned}
        &\sigma_i^+={\rm e}^{{\rm i\pi}\sum_{k=1}^{i-1}\sigma^+_k\sigma^-_k}c_i^{\dagger},\\
        &\sigma_i^-={\rm e}^{{\rm -i\pi}\sum_{k=1}^{i-1}\sigma^+_k\sigma^-_k}c_i,
    \end{aligned}
\end{equation}
where the exponentials are called string operators to reconcile the anti-commute relation for fermionic operators and the commute relation for spin operators on different sites. 
From the transformation, we can easily derive,
\begin{equation}
    \label{2301121100}
    \begin{aligned}
        \sigma_i^+\sigma_i^-=c_i^{\dagger}c_i=n_i,
    \end{aligned}
\end{equation}
where $n_i$ is the occupation operator on the site $i$. Moreover, for Pauli $Z$ operator, we have,
\begin{equation}
    \label{2301121103}
    \begin{aligned}
        Z_i&=2\sigma_i^+\sigma_i^--1\\
        &=2c_i^{\dagger}c_i-1\\
        &=2n_i-1.
    \end{aligned}
\end{equation}
From the above relations, we can rewrite string operators as,
\begin{equation}
    \label{2301121129}
    \begin{aligned}
        &{\rm e}^{{\rm \pm i\pi}\sum_{k=1}^{i-1}\sigma^+_k\sigma^-_k}\\
        &=\prod_{k=1}^{i-1}{\rm e}^{{\rm \pm i\pi}\sigma_k^+\sigma_k^-}\\
        &=\prod_{k=1}^{i-1}\left({\rm e}^0\ket{0}\bra{0}+{\rm e}^{\rm \pm i\pi}\ket{1}\bra{1}\right)\\
        &=\prod_{k=1}^{i-1}\left(-Z\right)\\
        &=\prod_{k=1}^{i-1}\left(1-2n_k\right).
    \end{aligned}
\end{equation}
Consequently, Pauli $X$ operator can be written,
\begin{equation}
    \label{2301121645}
    \begin{aligned}
        X_i&=\sigma_i^+ + \sigma_i^-\\
        &=\prod_{k=1}^{i-1}\left(1-2n_k\right)\left(c_i+c_i^{\dagger}\right).
    \end{aligned}
\end{equation}\par

Next, we get the Hamiltonian Eq.~(\ref{2301142040}) in terms of the fermions operators using the above equations. For convenience, we operate a rotation between the X-axis and the Z-axis such that the form of the Hamiltonian is,
\begin{equation}
    \label{2301122007}
    \begin{aligned}
        H=-J_z\sum_{i=1}^{L}X_iX_{i+1}-h_1\sum_{i=1}^{L}\left(Z_i-X_{i-1}Z_iX_{i+1}\right).
    \end{aligned}
  \end{equation}
For the first term, we have,
\begin{equation}
    \label{2301122027}
    \begin{aligned}
        &X_iX_{i+1}\\
                &=\prod_{k=1}^{i-1}\left(1-2n_k\right)\left(c_i+c_i^{\dagger}\right)\prod_{k=1}^{i}\left(1-2n_k\right)\left(c_{i+1}+c_{i+1}^{\dagger}\right)\\
                &=\left[\prod_{k=1}^{i-1}\left(1-2n_k\right)\right]^2\left(c_i+c_i^{\dagger}\right)\left(1-2n_i\right)\left(c_{i+1}+c_{i+1}^{\dagger}\right)\\
                &=\left[\prod_{k=1}^{i-1}\left(-Z\right)\right]^2\left(c_i+c_i^{\dagger}\right)\left(1-2n_i\right)\left(c_{i+1}+c_{i+1}^{\dagger}\right)\\
                &=\left(c_i+c_i^{\dagger}\right)\left(1-2n_i\right)\left(c_{i+1}+c_{i+1}^{\dagger}\right)\\
                &=\left(c_i-2c_ic_i^{\dagger}c_i+c_i^{\dagger}-2c_i^{\dagger}c_i^{\dagger}c_i\right)\left(c_{i+1}+c^{\dagger}_{i+1}\right)\\
                &=\left[c_i-2c_i\left(1-c_ic_i^{\dagger}\right)+c_i^{\dagger}-2c_i^{\dagger}c_i^{\dagger}c_i\right]\left(c_{i+1}+c^{\dagger}_{i+1}\right)\\
                &=\left(c_i^{\dagger}-c_i\right)\left(c_{i+1}+c_{i+1}^{\dagger}\right),
    \end{aligned}
\end{equation}
where the properties $c^2=\left(c^{\dagger}\right)^2=0$ is used to get the last expression.
Then for the last term, we have,
\begin{equation}
    \label{2301131800}
    \begin{aligned}
        &X_{i-1}Z_iX_{i+1}\\
                &=\prod_{k=1}^{i-2}\left(1-2n_k\right)\left(c_{i-1}+c_{i-1}^{\dagger}\right)\left(2n_i-1\right)\prod_{k=1}^{i}\left(1-2n_k\right)\left(c_{i+1}+c_{i+1}^{\dagger}\right)\\
                &=\left(c_{i-1}+c_{i-1}^{\dagger}\right)\left(2n_i-1\right)\left(1-2n_{i-1}\right)\left(1-2n_i\right)\left(c_{i+1}+c_{i+1}^{\dagger}\right)\\
                &=\left(c_{i-1}+c_{i-1}^{\dagger}\right)\left(-1\right)\left(1-2n_{i-1}\right)\left(c_{i+1}+c_{i+1}^{\dagger}\right)\\
                &=\left(c_{i-1}-2c_{i-1}c_{i-1}^{\dagger}c_{i-1}+c_{i+1}^{\dagger}-2c_{i+1}^{\dagger}c_{i+1}^{\dagger}c_{i+1}\right)\left(-1\right)\left(c_{i+1}+c^{\dagger}_{i+1}\right)\\
                &=\left(c_{i-1}^{\dagger}-c_{i-1}\right)\left(-1\right)\left(c_{i+1}+c_{i+1}^{\dagger}\right).
    \end{aligned}
\end{equation}
The combination of Eq.~(\ref{2301121103}), Eq.~(\ref{2301122027}), and Eq.~(\ref{2301131800}) gives,
\begin{equation}
    \label{2301122207}
    \begin{aligned}
        H=&-J_z\sum_{i=1}^{L}\left(c_i^{\dagger}-c_i\right)\left(c_{i+1}+c_{i+1}^{\dagger}\right)\\
          &-h_1\sum_{i=1}^{L}\left(2c_i^{\dagger}c_i-1\right)\\
          &-h_1\sum_{i=1}^{L}\left(c_{i-1}^{\dagger}-c_{i-1}\right)\left(c_{i+1}+c_{i+1}^{\dagger}\right).
    \end{aligned}
\end{equation}\par
\begin{figure*}[hp]
    \centering
    \includegraphics[width=0.5\textwidth]{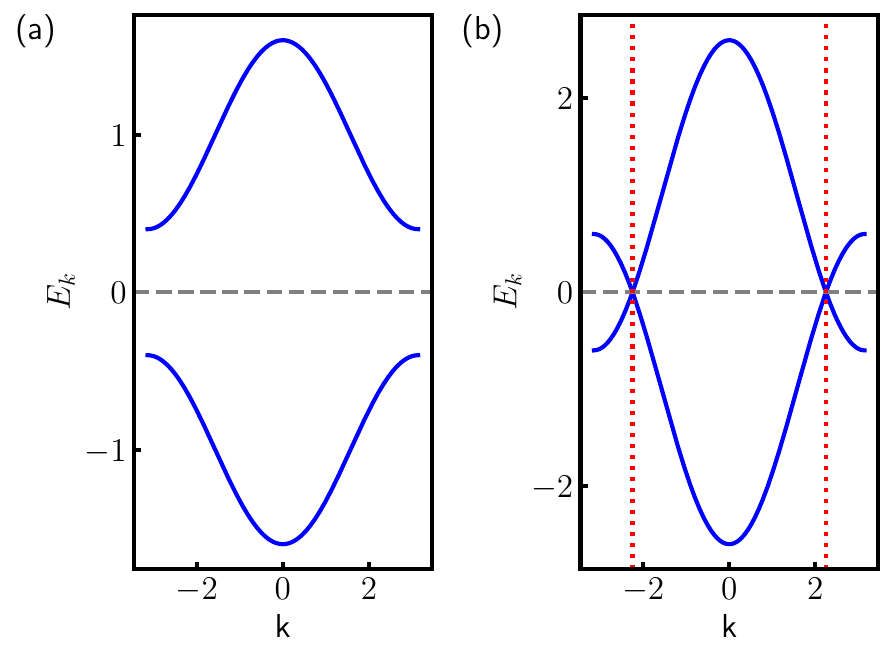}
    \caption{Dispersion relation of the model~\eqref{2301142040}. 
             (a)~For $J_z=1$ and $h_1=0.3$, one has $h_1/J_z<\frac12$, and the dispersion relation shows that the system is gapped.
             (b)~For $J_z=1$ and $h_1=0.8$, one finds $h_1/J_z>\frac12$, and the dispersion relation indicates a gapless phase; Dirac cones appear at $k={\pm \arccos}\bigl(-\frac{J_z}{2h_1}\bigr)$, marked by the red dashed lines.}
    \label{fig:dispersion}
\end{figure*}

We now turn to calculating the dispersion relation of Eq. (\ref{2301122207}). Considering the translation invariance of the system, and setting the lattice constant of the model $a=1$, we take the Fourier transform of Eq. (\ref{2301122207}) as follows,
for the first term,
\begin{equation}
    \label{2301131852}
    \begin{aligned}
        &-J_z\sum_{i=1}^{L}\left(c_i^{\dagger}-c_i\right)\left(c_{i+1}+c_{i+1}^{\dagger}\right)\\
        &=-J_z\frac{1}{L}\sum_{k, k'}\sum_{i=1}^{L}\left(c_{-k}^{\dagger}-c_k\right){\rm e}^{{\rm i}kx_i}\left(c_{k'}+c_{-k'}^{\dagger}\right){\rm e}^{{\rm i}k'\left(x_i+1\right)}\\
                    &=-J_z\frac{1}{L}\sum_{k, k'}\sum_{i=1}^{L}\left(c_{-k}^{\dagger}-c_k\right)\left(c_{k'}+c_{-k'}^{\dagger}\right){\rm e}^{{\rm i}x_i\left(k+k'\right)}{\rm e}^{{\rm i}k'}\\
                    &=-J_z\sum_{k, k'}\left(c_{-k}^{\dagger}-c_k\right)\left(c_{k'}+c_{-k'}^{\dagger}\right)\delta\left(k+k'=0\right){\rm e}^{{\rm i}k'}\\
                    &=-J_z\sum_{k}\left(c_{-k}^{\dagger}-c_k\right)\left(c_{-k}+c_{k}^{\dagger}\right){\rm e}^{{\rm -i}k}\\
                    &=-\frac{J_z}{2}\sum_{k}\left[\left(c_{-k}^{\dagger}-c_k\right)\left(c_{-k}+c_{k}^{\dagger}\right){\rm e}^{{\rm -i}k}+\left(c_{k}^{\dagger}-c_{-k}\right)\left(c_{k}+c_{-k}^{\dagger}\right){\rm e}^{{\rm i}k}\right]\\
                    &=-J_z\sum_{k}\left[\left(c_{k}^{\dagger}c_{-k}^{\dagger}-c_{-k}c_k\right)\frac{{\rm e}^{{\rm i}k}-{\rm e}^{{\rm -i}k}}{2}+\left(-c_{k}c_{k}^{\dagger}{\rm e}^{{\rm -i}k}+c_{k}^{\dagger}c_{k}{\rm e}^{{\rm i}k}\right)\right]\\
                    &=-J_z\sum_{k}\left[\left(c_{k}^{\dagger}c_{-k}^{\dagger}-c_{-k}c_k\right){\rm isin}k+2c_{k}^{\dagger}c_{k}{\rm cos}k\right]\\
                    &=-J_z\sum_{k}
                    \begin{pmatrix}
                        c_{k}^{\dagger}&c_{-k}
                    \end{pmatrix}
                    \begin{pmatrix}
                        {\rm cos}k&{\rm isin}k\\
                        {\rm -isin}k&{\rm -cos}k
                    \end{pmatrix}
                    \begin{pmatrix}
                        c_{k}\\
                        c_{-k}^{\dagger}
                    \end{pmatrix},
    \end{aligned}
\end{equation}
for the second term,
\begin{equation}
    \label{2301132045}
    \begin{aligned}
        &-h_1\sum_{i=1}^{L}\left(2c_i^{\dagger}c_i-1\right)\\
                    &=-h_1\frac{1}{L}\sum_{k, k'}\sum_{i=1}^{L}\left(2c_{-k}^{\dagger}c_{k'}\right){\rm e}^{{\rm i}kx_i}{\rm e}^{{\rm i}k'x_i}+const.\\
                    &=-h_1\sum_{k, k'}2c_{-k}^{\dagger}c_{k'}\delta\left(k+k'=0\right)\\
                    &=-h_1\sum_{k}2c_{k}^{\dagger}c_k\\
                    &=-h_1\sum_{k}
                    \begin{pmatrix}
                        c_{k}^{\dagger}&c_{-k}
                    \end{pmatrix}
                    \begin{pmatrix}
                        1&0\\
                        0&-1
                    \end{pmatrix}
                    \begin{pmatrix}
                        c_{k}\\
                        c_{-k}^{\dagger}
                    \end{pmatrix},
    \end{aligned}
\end{equation}
and for the third term,
\begin{equation}
    \label{2301132129}
    \begin{aligned}
        &-h_1\sum_{i=1}^{L}\left(c_{i-1}^{\dagger}-c_{i-1}\right)\left(c_{i+1}+c_{i+1}^{\dagger}\right)\\
        &=-h_1\sum_{k}
        \begin{pmatrix}
            c_{k}^{\dagger}&c_{-k}
        \end{pmatrix}
        \begin{pmatrix}
            {\rm cos}2k&{\rm isin}2k\\
            {\rm -isin}2k&{\rm -cos}2k
        \end{pmatrix}
        \begin{pmatrix}
            c_{k}\\
            c_{-k}^{\dagger}
        \end{pmatrix}.
    \end{aligned}
\end{equation}
Adding the terms Eq. (\ref{2301131852}), Eq. (\ref{2301132045}), and Eq. (\ref{2301132129}) together, we have the Hamiltonian of the model Eq. (\ref{2301122207}) in k-space,
\begin{widetext}
\begin{equation}
    \label{2301132133}
    \begin{aligned}
        H &=\sum_{k}
        \begin{pmatrix}
            c_{k}^{\dagger}&c_{-k}
        \end{pmatrix}
        \begin{pmatrix}
            -J_z{\rm cos}k-h_1\left({\rm cos}2k+1\right)&{\rm -i}J_z{\rm sin}k-{\rm i}h_1{\rm sin}2k\\
            {\rm i}J_z{\rm sin}k+{\rm i}h_1{\rm sin}2k&J_z{\rm cos}k+h_1\left({\rm cos}2k+1\right)
        \end{pmatrix}
        \begin{pmatrix}
            c_{k}\\
            c_{-k}^{\dagger}
        \end{pmatrix}\\
        &=\sum_{k}
        \begin{pmatrix}
            c_{k}^{\dagger}&c_{-k}
        \end{pmatrix}
        \begin{pmatrix}
            -{\rm cos}k\left(J_z+2h_1{\rm cos}k\right)&{\rm -i}{\rm sin}k\left(J_z+2h_1{\rm cos}k\right)\\
            {\rm i}{\rm sin}k\left(J_z+2h_1{\rm cos}k\right)&{\rm cos}k\left(J_z+2h_1{\rm cos}2k\right)
        \end{pmatrix}
        \begin{pmatrix}
            c_{k}\\
            c_{-k}^{\dagger}
        \end{pmatrix},    
    \end{aligned}
\end{equation}
\end{widetext}
where we use the double-angle formula for sine and cosine. Then we can obtain the eigenvalue $E_k$ by diagonalization of the matrix,
\begin{equation}
    \label{2301132209}
    \begin{aligned}
        E_k=\pm\left(J_z+2h_1{\rm cos}k\right).
    \end{aligned}
\end{equation}
We then plot the dispersion relation, as shown in Fig.~\ref{fig:dispersion}.

According to the energy spectrum~\eqref{2301132209} and the Fig.~\ref{fig:dispersion}, we can see that the model has two Dirac cones with $E_k=0$ at $k={\rm \pm arccos}\left(-\frac{J_z}{2h_1}\right)$. 
In addition, note that the Dirac cones are always present in the parameter interval $J_z<2h_1$, which is different with Ising model with anomaly-free $\mathbb Z_2$ symmetry. 
Considering the central charge $c=\frac{1}{2}$ for one Dirac cone, and the Luttinger parameter $K=1$ for free fermion system, the model can be described by $c=1$ CFT with $K=1$.

\bibliography{ref}

\begin{thebibliography}{26}%
\makeatletter
\providecommand \@ifxundefined [1]{%
 \@ifx{#1\undefined}
}%
\providecommand \@ifnum [1]{%
 \ifnum #1\expandafter \@firstoftwo
 \else \expandafter \@secondoftwo
 \fi
}%
\providecommand \@ifx [1]{%
 \ifx #1\expandafter \@firstoftwo
 \else \expandafter \@secondoftwo
 \fi
}%
\providecommand \natexlab [1]{#1}%
\providecommand \enquote  [1]{``#1''}%
\providecommand \bibnamefont  [1]{#1}%
\providecommand \bibfnamefont [1]{#1}%
\providecommand \citenamefont [1]{#1}%
\providecommand \href@noop [0]{\@secondoftwo}%
\providecommand \href [0]{\begingroup \@sanitize@url \@href}%
\providecommand \@href[1]{\@@startlink{#1}\@@href}%
\providecommand \@@href[1]{\endgroup#1\@@endlink}%
\providecommand \@sanitize@url [0]{\catcode `\\12\catcode `\$12\catcode `\&12\catcode `\#12\catcode `\^12\catcode `\_12\catcode `\%12\relax}%
\providecommand \@@startlink[1]{}%
\providecommand \@@endlink[0]{}%
\providecommand \url  [0]{\begingroup\@sanitize@url \@url }%
\providecommand \@url [1]{\endgroup\@href {#1}{\urlprefix }}%
\providecommand \urlprefix  [0]{URL }%
\providecommand \Eprint [0]{\href }%
\providecommand \doibase [0]{https://doi.org/}%
\providecommand \selectlanguage [0]{\@gobble}%
\providecommand \bibinfo  [0]{\@secondoftwo}%
\providecommand \bibfield  [0]{\@secondoftwo}%
\providecommand \translation [1]{[#1]}%
\providecommand \BibitemOpen [0]{}%
\providecommand \bibitemStop [0]{}%
\providecommand \bibitemNoStop [0]{.\EOS\space}%
\providecommand \EOS [0]{\spacefactor3000\relax}%
\providecommand \BibitemShut  [1]{\csname bibitem#1\endcsname}%
\let\auto@bib@innerbib\@empty
\bibitem [{\citenamefont {Landau}(1937)}]{landau1937theorie}%
  \BibitemOpen
  \bibfield  {author} {\bibinfo {author} {\bibfnamefont {L.~D.}\ \bibnamefont {Landau}},\ }\bibfield  {title} {\bibinfo {title} {Zur theorie der phasenumwandlungen ii},\ }\href@noop {} {\bibfield  {journal} {\bibinfo  {journal} {Phys. Z. Sowjetunion}\ }\textbf {\bibinfo {volume} {11}},\ \bibinfo {pages} {26} (\bibinfo {year} {1937})}\BibitemShut {NoStop}%
\bibitem [{\citenamefont {Wen}\ and\ \citenamefont {Niu}(1990)}]{WenNiu1990TO}%
  \BibitemOpen
  \bibfield  {author} {\bibinfo {author} {\bibfnamefont {X.~G.}\ \bibnamefont {Wen}}\ and\ \bibinfo {author} {\bibfnamefont {Q.}~\bibnamefont {Niu}},\ }\bibfield  {title} {\bibinfo {title} {Ground-state degeneracy of the fractional quantum hall states in the presence of a random potential and on high-genus riemann surfaces},\ }\href {https://doi.org/10.1103/PhysRevB.41.9377} {\bibfield  {journal} {\bibinfo  {journal} {Phys. Rev. B}\ }\textbf {\bibinfo {volume} {41}},\ \bibinfo {pages} {9377} (\bibinfo {year} {1990})}\BibitemShut {NoStop}%
\bibitem [{\citenamefont {Wen}(1990)}]{Wen1990TO}%
  \BibitemOpen
  \bibfield  {author} {\bibinfo {author} {\bibfnamefont {X.~G.}\ \bibnamefont {Wen}},\ }\bibfield  {title} {\bibinfo {title} {Topological orders in rigid states},\ }\href {https://doi.org/10.1142/S0217979290000139} {\bibfield  {journal} {\bibinfo  {journal} {Int. J. Mod. Phys. B}\ }\textbf {\bibinfo {volume} {04}},\ \bibinfo {pages} {239} (\bibinfo {year} {1990})}\BibitemShut {NoStop}%
\bibitem [{\citenamefont {Gu}\ and\ \citenamefont {Wen}(2009)}]{Gu2009}%
  \BibitemOpen
  \bibfield  {author} {\bibinfo {author} {\bibfnamefont {Z.-C.}\ \bibnamefont {Gu}}\ and\ \bibinfo {author} {\bibfnamefont {X.-G.}\ \bibnamefont {Wen}},\ }\bibfield  {title} {\bibinfo {title} {Tensor-entanglement-filtering renormalization approach and symmetry-protected topological order},\ }\href {https://doi.org/10.1103/PhysRevB.80.155131} {\bibfield  {journal} {\bibinfo  {journal} {Phys. Rev. B}\ }\textbf {\bibinfo {volume} {80}},\ \bibinfo {pages} {155131} (\bibinfo {year} {2009})}\BibitemShut {NoStop}%
\bibitem [{\citenamefont {Chen}\ \emph {et~al.}(2011)\citenamefont {Chen}, \citenamefont {Liu},\ and\ \citenamefont {Wen}}]{ChenCZX}%
  \BibitemOpen
  \bibfield  {author} {\bibinfo {author} {\bibfnamefont {X.}~\bibnamefont {Chen}}, \bibinfo {author} {\bibfnamefont {Z.-X.}\ \bibnamefont {Liu}},\ and\ \bibinfo {author} {\bibfnamefont {X.-G.}\ \bibnamefont {Wen}},\ }\bibfield  {title} {\bibinfo {title} {Two-dimensional symmetry-protected topological orders and their protected gapless edge excitations},\ }\href {https://doi.org/10.1103/PhysRevB.84.235141} {\bibfield  {journal} {\bibinfo  {journal} {Phys. Rev. B}\ }\textbf {\bibinfo {volume} {84}},\ \bibinfo {pages} {235141} (\bibinfo {year} {2011})}\BibitemShut {NoStop}%
\bibitem [{\citenamefont {Chen}\ \emph {et~al.}(2012)\citenamefont {Chen}, \citenamefont {Gu}, \citenamefont {Liu},\ and\ \citenamefont {Wen}}]{ChenSPTScience}%
  \BibitemOpen
  \bibfield  {author} {\bibinfo {author} {\bibfnamefont {X.}~\bibnamefont {Chen}}, \bibinfo {author} {\bibfnamefont {Z.-C.}\ \bibnamefont {Gu}}, \bibinfo {author} {\bibfnamefont {Z.-X.}\ \bibnamefont {Liu}},\ and\ \bibinfo {author} {\bibfnamefont {X.-G.}\ \bibnamefont {Wen}},\ }\bibfield  {title} {\bibinfo {title} {Symmetry-protected topological orders in interacting bosonic systems},\ }\href {https://doi.org/10.1126/science.1227224} {\bibfield  {journal} {\bibinfo  {journal} {Science}\ }\textbf {\bibinfo {volume} {338}},\ \bibinfo {pages} {1604} (\bibinfo {year} {2012})}\BibitemShut {NoStop}%
\bibitem [{\citenamefont {Chen}\ \emph {et~al.}(2013)\citenamefont {Chen}, \citenamefont {Gu}, \citenamefont {Liu},\ and\ \citenamefont {Wen}}]{ChenSPTPRB}%
  \BibitemOpen
  \bibfield  {author} {\bibinfo {author} {\bibfnamefont {X.}~\bibnamefont {Chen}}, \bibinfo {author} {\bibfnamefont {Z.-C.}\ \bibnamefont {Gu}}, \bibinfo {author} {\bibfnamefont {Z.-X.}\ \bibnamefont {Liu}},\ and\ \bibinfo {author} {\bibfnamefont {X.-G.}\ \bibnamefont {Wen}},\ }\bibfield  {title} {\bibinfo {title} {Symmetry protected topological orders and the group cohomology of their symmetry group},\ }\href {https://doi.org/10.1103/PhysRevB.87.155114} {\bibfield  {journal} {\bibinfo  {journal} {Phys. Rev. B}\ }\textbf {\bibinfo {volume} {87}},\ \bibinfo {pages} {155114} (\bibinfo {year} {2013})}\BibitemShut {NoStop}%
\bibitem [{\citenamefont {Hasan}\ and\ \citenamefont {Kane}(2010)}]{Hasan2010}%
  \BibitemOpen
  \bibfield  {author} {\bibinfo {author} {\bibfnamefont {M.~Z.}\ \bibnamefont {Hasan}}\ and\ \bibinfo {author} {\bibfnamefont {C.~L.}\ \bibnamefont {Kane}},\ }\bibfield  {title} {\bibinfo {title} {\textit{Colloquium} : Topological insulators},\ }\href {https://doi.org/10.1103/RevModPhys.82.3045} {\bibfield  {journal} {\bibinfo  {journal} {Rev. Mod. Phys.}\ }\textbf {\bibinfo {volume} {82}},\ \bibinfo {pages} {3045} (\bibinfo {year} {2010})}\BibitemShut {NoStop}%
\bibitem [{\citenamefont {Qi}\ and\ \citenamefont {Zhang}(2011)}]{Qi2011}%
  \BibitemOpen
  \bibfield  {author} {\bibinfo {author} {\bibfnamefont {X.-L.}\ \bibnamefont {Qi}}\ and\ \bibinfo {author} {\bibfnamefont {S.-C.}\ \bibnamefont {Zhang}},\ }\bibfield  {title} {\bibinfo {title} {Topological insulators and superconductors},\ }\href {https://doi.org/10.1103/RevModPhys.83.1057} {\bibfield  {journal} {\bibinfo  {journal} {Rev. Mod. Phys.}\ }\textbf {\bibinfo {volume} {83}},\ \bibinfo {pages} {1057} (\bibinfo {year} {2011})}\BibitemShut {NoStop}%
\bibitem [{\citenamefont {Wen}(2013)}]{Wen2013Anomaly}%
  \BibitemOpen
  \bibfield  {author} {\bibinfo {author} {\bibfnamefont {X.-G.}\ \bibnamefont {Wen}},\ }\bibfield  {title} {\bibinfo {title} {Classifying gauge anomalies through symmetry-protected trivial orders and classifying gravitational anomalies through topological orders},\ }\href {https://doi.org/10.1103/PhysRevD.88.045013} {\bibfield  {journal} {\bibinfo  {journal} {Phys. Rev. D}\ }\textbf {\bibinfo {volume} {88}},\ \bibinfo {pages} {045013} (\bibinfo {year} {2013})}\BibitemShut {NoStop}%
\bibitem [{\citenamefont {Cheng}\ \emph {et~al.}(2016)\citenamefont {Cheng}, \citenamefont {Zaletel}, \citenamefont {Barkeshli}, \citenamefont {Vishwanath},\ and\ \citenamefont {Bonderson}}]{Meng2016LSM}%
  \BibitemOpen
  \bibfield  {author} {\bibinfo {author} {\bibfnamefont {M.}~\bibnamefont {Cheng}}, \bibinfo {author} {\bibfnamefont {M.}~\bibnamefont {Zaletel}}, \bibinfo {author} {\bibfnamefont {M.}~\bibnamefont {Barkeshli}}, \bibinfo {author} {\bibfnamefont {A.}~\bibnamefont {Vishwanath}},\ and\ \bibinfo {author} {\bibfnamefont {P.}~\bibnamefont {Bonderson}},\ }\bibfield  {title} {\bibinfo {title} {Translational symmetry and microscopic constraints on symmetry-enriched topological phases: A view from the surface},\ }\href {https://doi.org/10.1103/PhysRevX.6.041068} {\bibfield  {journal} {\bibinfo  {journal} {Phys. Rev. X}\ }\textbf {\bibinfo {volume} {6}},\ \bibinfo {pages} {041068} (\bibinfo {year} {2016})}\BibitemShut {NoStop}%
\bibitem [{\citenamefont {Wang}\ \emph {et~al.}()\citenamefont {Wang}, \citenamefont {Fang}, \citenamefont {Cheng}, \citenamefont {Qi},\ and\ \citenamefont {Meng}}]{YCWang2017QSL}%
  \BibitemOpen
  \bibfield  {author} {\bibinfo {author} {\bibfnamefont {Y.-C.}\ \bibnamefont {Wang}}, \bibinfo {author} {\bibfnamefont {C.}~\bibnamefont {Fang}}, \bibinfo {author} {\bibfnamefont {M.}~\bibnamefont {Cheng}}, \bibinfo {author} {\bibfnamefont {Y.}~\bibnamefont {Qi}},\ and\ \bibinfo {author} {\bibfnamefont {Z.~Y.}\ \bibnamefont {Meng}},\ }\bibfield  {title} {\bibinfo {title} {Topological spin liquid with symmetry-protected edge states},\ }\Eprint {https://arxiv.org/abs/1701.01552} {arXiv:1701.01552 [cond-mat.str-el]} \BibitemShut {NoStop}%
\bibitem [{\citenamefont {Drell}\ \emph {et~al.}(1976)\citenamefont {Drell}, \citenamefont {Weinstein},\ and\ \citenamefont {Yankielowicz}}]{SLACFermion}%
  \BibitemOpen
  \bibfield  {author} {\bibinfo {author} {\bibfnamefont {S.~D.}\ \bibnamefont {Drell}}, \bibinfo {author} {\bibfnamefont {M.}~\bibnamefont {Weinstein}},\ and\ \bibinfo {author} {\bibfnamefont {S.}~\bibnamefont {Yankielowicz}},\ }\bibfield  {title} {\bibinfo {title} {Strong-coupling field theories. ii. fermions and gauge fields on a lattice},\ }\href {https://doi.org/10.1103/PhysRevD.14.1627} {\bibfield  {journal} {\bibinfo  {journal} {Phys. Rev. D}\ }\textbf {\bibinfo {volume} {14}},\ \bibinfo {pages} {1627} (\bibinfo {year} {1976})}\BibitemShut {NoStop}%
\bibitem [{\citenamefont {Chatterjee}\ and\ \citenamefont {Wen}()}]{chatterjee2022algebra}%
  \BibitemOpen
  \bibfield  {author} {\bibinfo {author} {\bibfnamefont {A.}~\bibnamefont {Chatterjee}}\ and\ \bibinfo {author} {\bibfnamefont {X.-G.}\ \bibnamefont {Wen}},\ }\bibfield  {title} {\bibinfo {title} {Algebra of local symmetric operators and braided fusion $ n $-category--symmetry is a shadow of topological order},\ }\Eprint {https://arxiv.org/abs/2203.03596} {arXiv:2203.03596 [cond-mat.str-el]} \BibitemShut {NoStop}%
\bibitem [{\citenamefont {C{h}atterjee}\ and\ \citenamefont {Wen}()}]{chatterjee2022holographic}%
  \BibitemOpen
  \bibfield  {author} {\bibinfo {author} {\bibfnamefont {A.}~\bibnamefont {C{h}atterjee}}\ and\ \bibinfo {author} {\bibfnamefont {X.-G.}\ \bibnamefont {Wen}},\ }\bibfield  {title} {\bibinfo {title} {Holographic theory for the emergence and the symmetry protection of gaplessness and for continuous phase transitions},\ }\Eprint {https://arxiv.org/abs/2205.06244} {arXiv:2205.06244 [cond-mat.str-el]} \BibitemShut {NoStop}%
\bibitem [{\citenamefont {Ji}\ and\ \citenamefont {Wen}(2020)}]{ji2020categorical}%
  \BibitemOpen
  \bibfield  {author} {\bibinfo {author} {\bibfnamefont {W.}~\bibnamefont {Ji}}\ and\ \bibinfo {author} {\bibfnamefont {X.-G.}\ \bibnamefont {Wen}},\ }\bibfield  {title} {\bibinfo {title} {Categorical symmetry and noninvertible anomaly in symmetry-breaking and topological phase transitions},\ }\href {https://doi.org/10.1103/PhysRevResearch.2.033417} {\bibfield  {journal} {\bibinfo  {journal} {Phys. Rev. Research}\ }\textbf {\bibinfo {volume} {2}},\ \bibinfo {pages} {033417} (\bibinfo {year} {2020})}\BibitemShut {NoStop}%
\bibitem [{\citenamefont {Levin}\ and\ \citenamefont {Gu}(2012)}]{LevinGu2012}%
  \BibitemOpen
  \bibfield  {author} {\bibinfo {author} {\bibfnamefont {M.}~\bibnamefont {Levin}}\ and\ \bibinfo {author} {\bibfnamefont {Z.-C.}\ \bibnamefont {Gu}},\ }\bibfield  {title} {\bibinfo {title} {Braiding statistics approach to symmetry-protected topological phases},\ }\href {https://doi.org/10.1103/PhysRevB.86.115109} {\bibfield  {journal} {\bibinfo  {journal} {Phys. Rev. B}\ }\textbf {\bibinfo {volume} {86}},\ \bibinfo {pages} {115109} (\bibinfo {year} {2012})}\BibitemShut {NoStop}%
\bibitem [{\citenamefont {Kong}\ and\ \citenamefont {Zheng}(2020)}]{Kong2020}%
  \BibitemOpen
  \bibfield  {author} {\bibinfo {author} {\bibfnamefont {L.}~\bibnamefont {Kong}}\ and\ \bibinfo {author} {\bibfnamefont {H.}~\bibnamefont {Zheng}},\ }\bibfield  {title} {\bibinfo {title} {A mathematical theory of gapless edges of 2d topological orders. part {I}},\ }\href {https://doi.org/10.1007/JHEP02(2020)150} {\bibfield  {journal} {\bibinfo  {journal} {J. High Energy Phys.}\ }\textbf {\bibinfo {volume} {2020}}\bibinfo  {number} { (2)},\ \bibinfo {pages} {150}}\BibitemShut {NoStop}%
\bibitem [{\citenamefont {Kong}\ and\ \citenamefont {Zheng}(2021)}]{Kong2021}%
  \BibitemOpen
\bibfield  {number} {  }\bibfield  {author} {\bibinfo {author} {\bibfnamefont {L.}~\bibnamefont {Kong}}\ and\ \bibinfo {author} {\bibfnamefont {H.}~\bibnamefont {Zheng}},\ }\bibfield  {title} {\bibinfo {title} {A mathematical theory of gapless edges of 2d topological orders. part {II}},\ }\href {https://doi.org/https://doi.org/10.1016/j.nuclphysb.2021.115384} {\bibfield  {journal} {\bibinfo  {journal} {Nucl. Phys. B}\ }\textbf {\bibinfo {volume} {966}},\ \bibinfo {pages} {115384} (\bibinfo {year} {2021})}\BibitemShut {NoStop}%
\bibitem [{\citenamefont {Scaffidi}\ and\ \citenamefont {Ringel}(2016)}]{Scaffidi2016}%
  \BibitemOpen
  \bibfield  {author} {\bibinfo {author} {\bibfnamefont {T.}~\bibnamefont {Scaffidi}}\ and\ \bibinfo {author} {\bibfnamefont {Z.}~\bibnamefont {Ringel}},\ }\bibfield  {title} {\bibinfo {title} {Wave functions of symmetry-protected topological phases from conformal field theories},\ }\href {https://doi.org/10.1103/PhysRevB.93.115105} {\bibfield  {journal} {\bibinfo  {journal} {Phys. Rev. B}\ }\textbf {\bibinfo {volume} {93}},\ \bibinfo {pages} {115105} (\bibinfo {year} {2016})}\BibitemShut {NoStop}%
\bibitem [{\citenamefont {Cheng}\ and\ \citenamefont {Williamson}(2020)}]{Meng2020RA}%
  \BibitemOpen
  \bibfield  {author} {\bibinfo {author} {\bibfnamefont {M.}~\bibnamefont {Cheng}}\ and\ \bibinfo {author} {\bibfnamefont {D.~J.}\ \bibnamefont {Williamson}},\ }\bibfield  {title} {\bibinfo {title} {Relative anomaly in ($1+1$)d rational conformal field theory},\ }\href {https://doi.org/10.1103/PhysRevResearch.2.043044} {\bibfield  {journal} {\bibinfo  {journal} {Phys. Rev. Research}\ }\textbf {\bibinfo {volume} {2}},\ \bibinfo {pages} {043044} (\bibinfo {year} {2020})}\BibitemShut {NoStop}%
\bibitem [{\citenamefont {Fishman}\ \emph {et~al.}()\citenamefont {Fishman}, \citenamefont {White},\ and\ \citenamefont {Stoudenmire}}]{itensor}%
  \BibitemOpen
  \bibfield  {author} {\bibinfo {author} {\bibfnamefont {M.}~\bibnamefont {Fishman}}, \bibinfo {author} {\bibfnamefont {S.~R.}\ \bibnamefont {White}},\ and\ \bibinfo {author} {\bibfnamefont {E.~M.}\ \bibnamefont {Stoudenmire}},\ }\href@noop {} {\bibinfo {title} {The \mbox{ITensor} software library for tensor network calculations}},\ \Eprint {https://arxiv.org/abs/2007.14822} {arXiv:2007.14822 [cs.MS]} \BibitemShut {NoStop}%
\bibitem [{\citenamefont {Calabrese}\ and\ \citenamefont {Cardy}(2004)}]{calabrese2004entanglement}%
  \BibitemOpen
  \bibfield  {author} {\bibinfo {author} {\bibfnamefont {P.}~\bibnamefont {Calabrese}}\ and\ \bibinfo {author} {\bibfnamefont {J.}~\bibnamefont {Cardy}},\ }\bibfield  {title} {\bibinfo {title} {Entanglement entropy and quantum field theory},\ }\href {https://doi.org/https://doi.org/10.1088/1742-5468/2004/06/P06002} {\bibfield  {journal} {\bibinfo  {journal} {J. Stat. Mech.}\ }\textbf {\bibinfo {volume} {2004}},\ \bibinfo {pages} {P06002} (\bibinfo {year} {2004})}\BibitemShut {NoStop}%
\bibitem [{Note1()}]{Note1}%
  \BibitemOpen
  \bibinfo {note} {The peak is not divergent as $L$ increases, and this is consistent with the phase transition being of Berezinskii-Kosterlitz-Thouless type, which will be discussed later.}\BibitemShut {Stop}%
\bibitem [{\citenamefont {Luttinger}(1960)}]{luttinger1960fermi}%
  \BibitemOpen
  \bibfield  {author} {\bibinfo {author} {\bibfnamefont {J.}~\bibnamefont {Luttinger}},\ }\bibfield  {title} {\bibinfo {title} {Fermi surface and some simple equilibrium properties of a system of interacting fermions},\ }\href@noop {} {\bibfield  {journal} {\bibinfo  {journal} {Phys. Rev.}\ }\textbf {\bibinfo {volume} {119}},\ \bibinfo {pages} {1153} (\bibinfo {year} {1960})}\BibitemShut {NoStop}%
\bibitem [{Note2()}]{Note2}%
  \BibitemOpen
  \bibinfo {note} {For $h_2=0.6$, the critical point $h_{1c}\approx 1.16(4)$ is determined from the Binder ratio $U$ as a function of $h_1$ with different system sizes.}\BibitemShut {Stop}%
\end{thebibliography}%
\end{document}